\documentstyle[sprocl,epsfig]{article}

\def\be{\begin{equation}}
\def\ee{\end{equation}}
\def\bea{\begin{eqnarray}}
\def\eea{\end{eqnarray}}
\def\nn{\nonumber}
\def\als{\alpha_s}

\def\gev{\,{\rm GeV}}
\newcommand{\aem}{\alpha_{\rm em}}

\newcommand{\da}{{distribution amplitude}}
\newcommand{\das}{{distribution amplitudes}}
\newcommand{\wf}{wave function}
\newcommand{\wfs}{wave functions}
\newcommand{\spa}{soft physics approach}
\renewcommand{\d}{\rm d}
\newcommand{\ibid}[1]{{\it ibid.}~#1}

\newcommand{\AmS}{{\protect\the\textfont2
  A\kern-.1667em\lower.5ex\hbox{M}\kern-.125emS}}
\newcommand{\lsim}{\raisebox{-3pt}{$\,\stackrel{\textstyle <}{\sim}\,$}}

\begin{document}

\title{\begin{flushright}
               WU B 99-20 \\
               hep-ph/9908242\\[6mm]
       \end{flushright}
    SKEWED PARTON DISTRIBUTIONS AND REAL AND VIRTUAL COMPTON SCATTERING} 

\author{P. KROLL}

\address{Fachbereich Physik, Universit\"at Wuppertal\\ 
        Gau\ss strasse 20, D-42097 Wuppertal, Germany}%

\maketitle\abstracts{ The soft physics approach to Compton scattering
at moderately large momentum transfer is reviewed. It will be argued
that in that approach the Compton cross section as well as other
exclusive observables exhibit approximate scaling in a limited range
of momentum transfer.\\
Invited talk held at the workshop on exclusive and semi-exclusive
processes at high momentum transfer, Jefferson Lab, May (1999).}

\section{Introduction}

\begin{figure}
\parbox{\textwidth}{\begin{center}
   \psfig{file=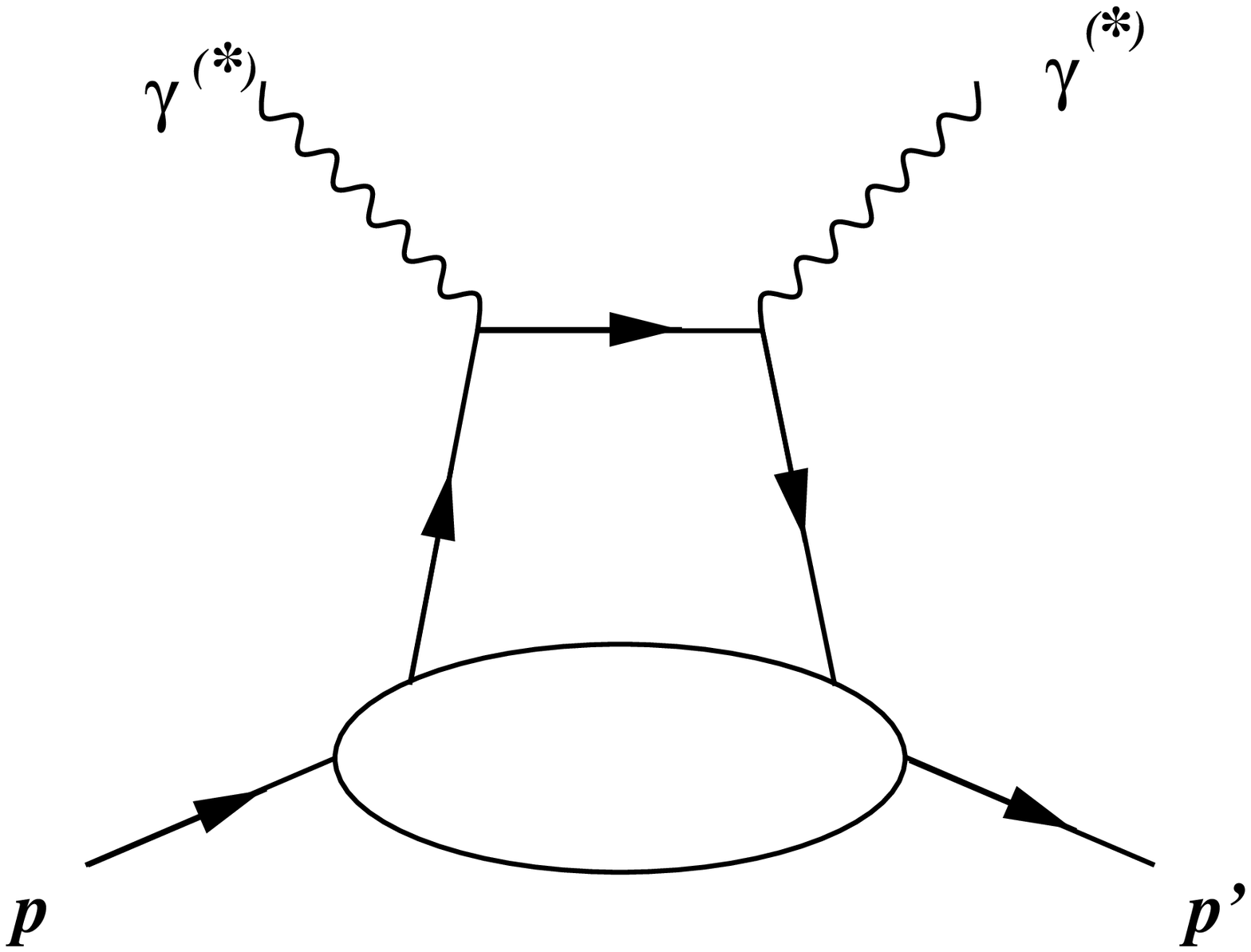,%
          bbllx=45pt,bblly=230pt,bburx=545pt,bbury=610pt,%
           width=4.5cm,clip=}\hspace{0.5cm}
   \psfig{file=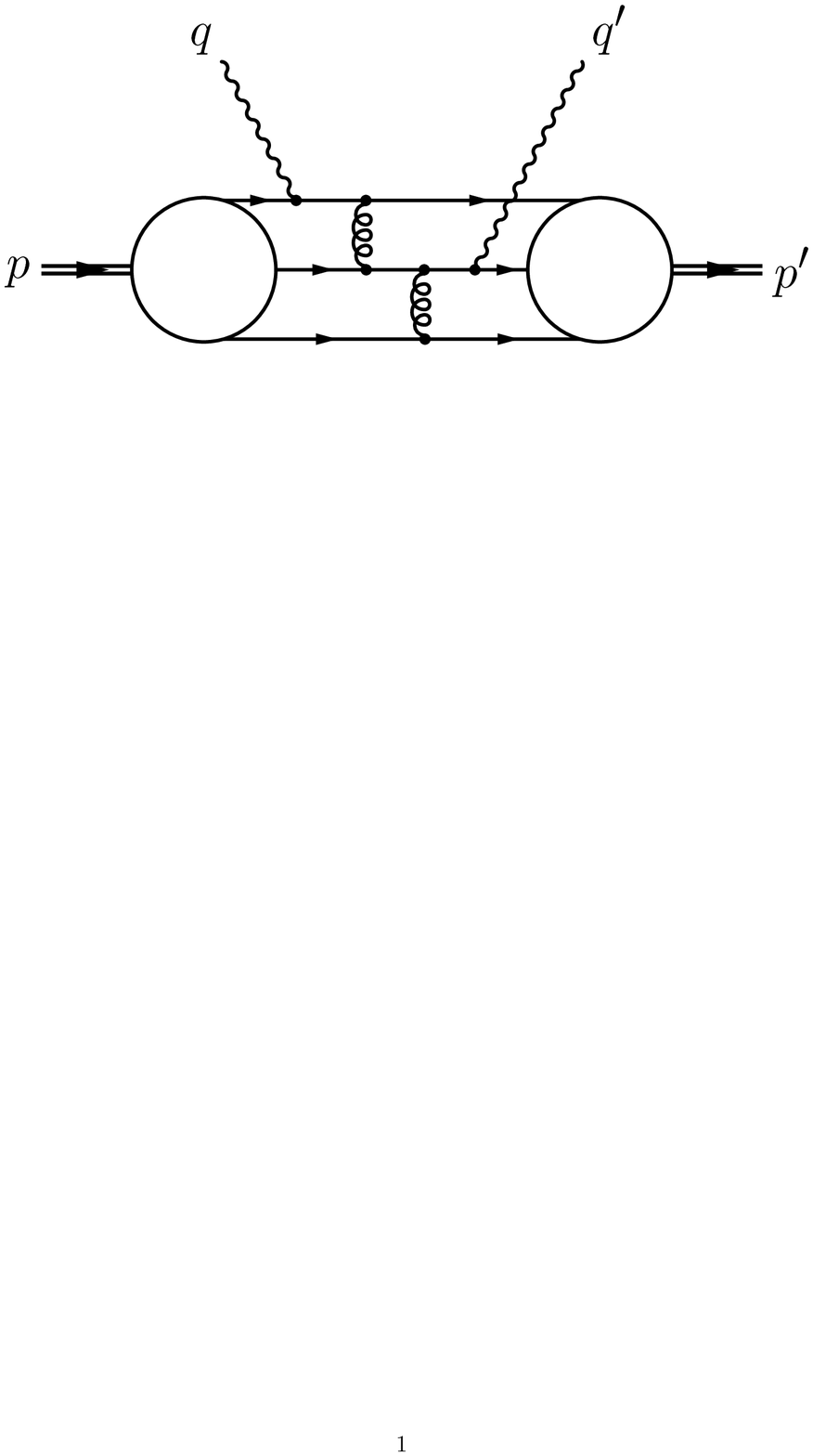,%
          bbllx=90pt,bblly=505pt,bburx=510pt,bbury=790pt,%
           width=4.5cm,clip=}
\end{center}}
\caption{The handbag diagram (left) and an example of a Feyman graph
         for Compton scattering within pQCD (right). The momenta of
         the incoming and outgoing protons (photons) are denoted by $p$ ($q$) 
         and $p'=p+\Delta$ ($q'=q-\Delta$), respectively. }
\label{fig:handbag}
\end{figure}

QCD provides three valence Fock state contributions to nucleon form
factors, real (RCS) and virtual (VCS) Compton scattering  at 
large momentum transfer: a soft overlap term with an active quark and two
spectators (see  Fig.\ \ref{fig:handbag}), the 
asymptotically dominant perturbative contribution where by means of
the exchange of two hard gluons the quarks are kept collinear
with respect to their parent nucleons (see Fig.\ \ref{fig:handbag}) and a
third contribution that is intermediate between the soft and the
perturbative contribution where only one hard gluon is exchanged and one of
the three quarks acts as a spectator. Both the soft and the intermediate terms
represent power corrections to the perturbative contribution. Higher
Fock state contributions are suppressed. The crucial
question is what is the relative strengths of the three
contributions at experimentally accessible values of momentum transfer,
i.e.\ at $-t$ of the order of 10 GeV$^2$? The pQCD followers (cf.\
Ref.\ \cite{niz91} for recent calculations of RCS) assume
the dominance of the perturbative contribution and neglect the other two
contributions while the soft physics community
presumes the dominance of the overlap contribution. Which group is
right is not yet fully decided although comparison with the pion case \cite{kro96}
seems to favour a strong overlap contribution.  

In this talk I am going to report on results for Compton scattering
obtained within the soft physics approach \cite{DFJK}. It has been shown
recently \cite{DFJK,rad98a} that at moderately large momentum
transfer Compton scattering off protons
approximately factorises into a hard photon-parton subprocess and soft
proton matrix elements described by new form factors specific to
Compton scattering. These form factors, as the ordinary
electromagnetic ones, represent moments of skewed parton distributions
(SPD) \cite{mue98}. The \spa{} bears resemblance to the fixed-pole
model advocated for in \cite{bro72} long time ago.

\section{The soft physics ap\-proach to Compton scattering}

For Mandelstam variables, $s$, $t$ and  $u$, that are large on a
hadronic scale the handbag diagram shown in Fig.\
\ref{fig:handbag} describes RCS and VCS. To see this it is of
advantage to choose a symmetric frame of reference where the plus and
minus light-cone components of $\Delta$ are zero.
This implies $t=-\Delta^2_\perp$ as well as a vanishing skewdness
parameter $\zeta=-\Delta^+/p^+$. To evaluate the SPD
appearing in the handbag diagram one may use a Fock state decomposition of the
proton and sum over all possible spectator
configurations. The crucial assumption is then that the soft hadron
\wfs{} are dominated by virtualities in the range $|k_i^2|\lsim
\Lambda^2$, where $\Lambda$ is a hadronic scale of the order of 1 GeV,
and by intrinsic transverse parton momenta, $k_{\perp i}$, defined
with respect to their parent hadron's momentum, that satisfy $k_{\perp
i}^2/x_i\lsim \Lambda^2$. Under this assumption factorisation of the
Compton amplitude in a hard photon-parton amplitude and $1/x$-moments
of SPDs is achieved \cite{DFJK}. 

As a consequence of this result the Compton amplitudes conserving the
proton helicity are given by
\be
{\cal M}_{\mu'+,\,\mu +} \,=\, \;2\pi\aem \left[{\cal
    H}_{\mu'+,\,\mu+}\,(R_V + R_A)
  \,+\, {\cal H}_{\mu'-,\,\mu-}\,(R_V - R_A) \right ]\,.
\label{final}
\ee
Proton helicity flip is neglected. $\mu$ and $\mu'$ are
the helicities of the incoming and outgoing photon in the
photon-proton cms, respectively. The photon-quark subprocess
amplitudes, ${\cal H}$, are calculated for massless quarks in lowest order QED.
The form factors in Eq.\ (\ref{final}), $R_V$ and
$R_A$, represent $1/x$-moments of SPDs at zero skewedness parameter.
$R_V$ is defined by
\begin{eqnarray} 
\sum_a e_a^2\, \int_0^1\, \frac{{\d} x}{x}\, p^+
   \int {{\d} z^-\over 2\pi}\, e^{i\, x p^+ z^-} 
     \langle p'|\,
     \overline\psi{}_{a}(0)\, \gamma^+\,\psi_{a}(z^-) - 
     \overline\psi{}_{a}(z^-)\, \gamma^+\,\psi_{a}(0) 
     \,| p \rangle  \nn \\[0.3em]
   = R_V(t)\, \bar{u}(p')\, \gamma^+ u(p)\, 
 + R_T(t)\, \frac{i}{2m}\bar{u}(p')
                          \sigma^{+\rho}\Delta_\rho u(p)\,,\hspace*{2cm}
\label{R-form-factors}
\eea
where the sum runs over quark flavours $a$ ($u$, $d$, \ldots), $e_a$
being the electric charge of quark $a$ in units of the positron
charge. $R_T$ being related to nucleon helicity flips, is neglected in
(\ref{final}). There is an analogous equation for the axial vector
nucleon matrix element, which defines the form factor $R_A$. 
Due to time reversal invariance the form factors $R_V$, $R_A$ etc.\
are real functions. 

As shown in \cite{DFJK} form factors can be represented as
generalized Drell-Yan light-cone \wf{} overlaps. Assuming a plausible
Gaussian  $k_{\perp i}$-depen\-dence of the soft Fock state \wfs{}, 
one can explicitly carry out the momentum integrations in
the Drell-Yan formula. For simplicity one may further assume a common
transverse size parameter, $\hat a$, for all Fock
states. This immediately allows one to sum over them, without
specifying the $x_i$-dependence of the \wf{}s. One then arrives at
\cite{DFJK,rad98a} 
\bea
F_1(t)&=& \sum_a\, e_a\, \int {\d} x\, 
         \exp{\left[\frac12 \hat a^2 t \frac{1-x}{x}\right]}    
                 \,\{ q_a(x) - \bar{q}_a(x) \} \,, \nn\\[0.5em]
R_V(t)&=& \sum_a\, e_a^2\, \int \frac {{\d} x}{x}\, 
         \exp{\left[\frac12 \hat a^2 t \frac{1-x}{x}\right]} 
                \,\{ q_a(x) + \bar{q}_a(x) \} \,,
\label{ffspd}
\eea
and the analogue for $R_A$ with $q_a+\bar{q}_a$ replaced by $\Delta
q_a + \Delta \bar{q}_a$. $q_a$ and $\Delta q_a$ are the usual
unpolarized and polarized parton distributions, respectively. 
The result for $F_1$ can also been found in \cite{bar93}. 

The only parameter appearing in (\ref{ffspd}) is the effective
transverse size parameter $\hat{a}$; it is known to be about 1
GeV{}$^{-1}$ with an uncertainty of about 20$\%$. Thus, this parameter
only allows some fine tuning of the results for the form factors.
Evaluating, for instance, the form factors from the parton
distributions derived by Gl\"uck et al. (GRV) \cite{GRV} with 
$\hat{a}=1\, GeV^{-1}$, one already finds good results.
Improvements are obtained by treating the lowest three Fock states
explicitly with specified $x$-dependencies. For the valence Fock
\da{}, for instance, a form proposed in Ref.~\cite{bol96} was used
\be 
 \Phi^{BK}_{123} \,=\, 60\, x_1x_2x_3\, (1+3x_1)\,,
\label{eq:BK}
\ee
valid at a factorisation scale of 1 GeV. All higher Fock states were
still treated in the global way, using the parton distributions of
Ref.~\cite{GRV} as input. 
Results for $t^2 F_1$ and $t^2 R_V$ obtained that way 
are displayed in Fig.\ \ref{fig:FFsoft}. Both the scaled
form factors, as well as $t^2 R_A$, exhibit broad maxima and,
hence, mimic dimensional counting rule behaviour in the $t$-range from 
about 5 to 15 GeV{}$^2$. For very large momentum transfer the form
factors turn gradually into the soft physics asymptotics $\sim
1/t^4$. This is the region where the perturbative contribution ($\sim
1/t^2$) takes the lead.  
\begin{figure}
\parbox{\textwidth}{\begin{center}
   \psfig{file=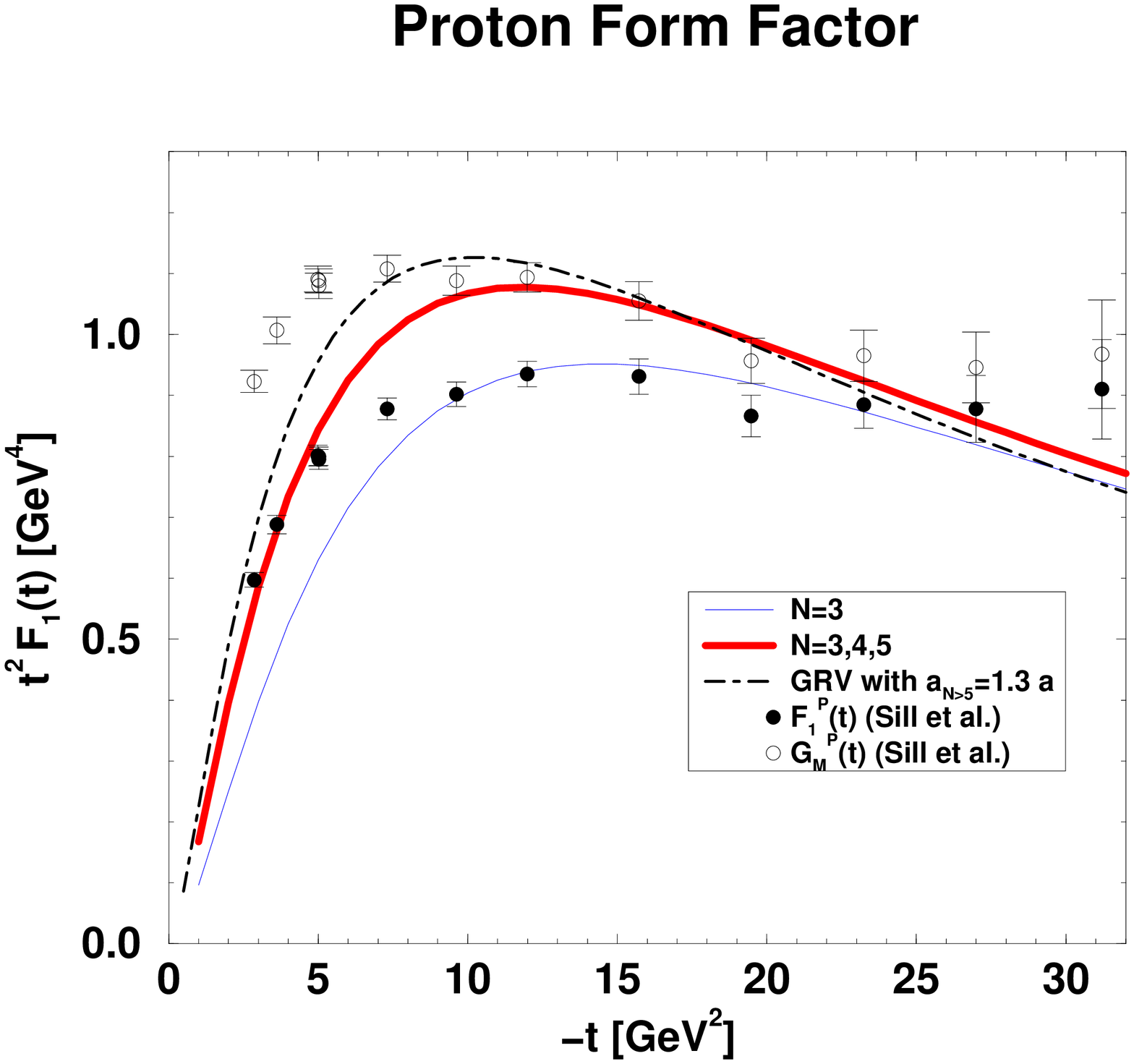, 
          bbllx=100pt,bblly=0pt,bburx=590pt,bbury=635pt,%
           width=4.3cm, angle=-90, clip=}\hspace{0.5cm}
      \psfig{file=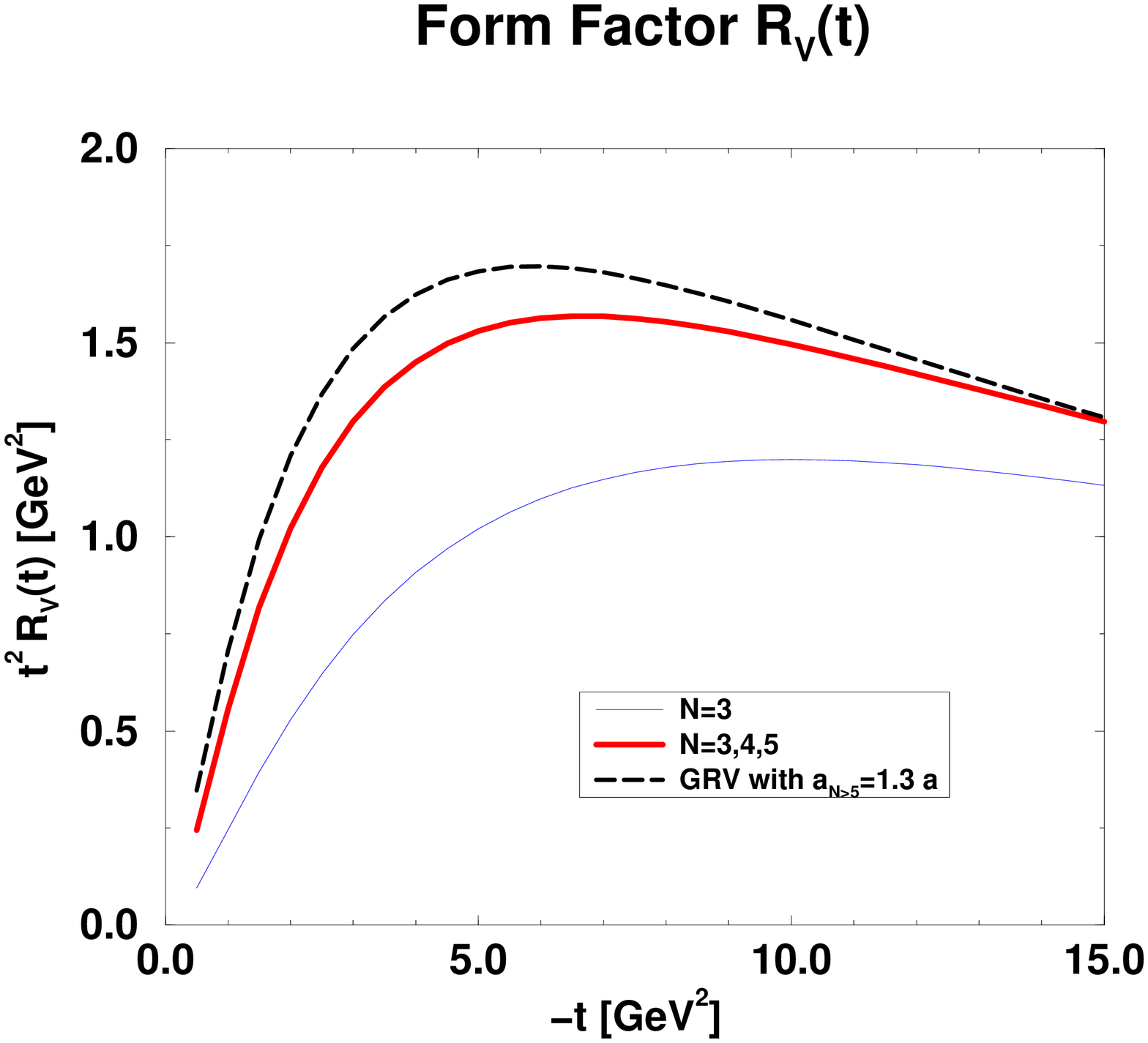, 
          bbllx=90pt,bblly=0pt,bburx=590pt,bbury=655pt,%
           width=4.3cm, angle=-90, clip=} 
\end{center}}
\caption{The Dirac (left) and the vector Compton (right) form factor
         of the proton as predicted by the soft physics approach 
         \protect\cite{DFJK,bol96}. Data are taken from
         \protect\cite{sil93}. The data on the magnetic form factor, $G_M$, are 
          shown in order to demonstrate the size of spin-flip effects.}
\label{fig:FFsoft}
\end{figure}

It is to be stressed that  the perturbative contribution to the proton
form factor evaluated from the \da{} (\ref{eq:BK}) within the modified perturbative
approach~\cite{bol96,ber95} is as small as only a few $\%$ of the  
experimental value. A similarly small value is expected for Compton
scattering. Therefore, the analysis performed in Ref.\ \cite{DFJK} is to be
regarded as a consistent calculation 
in which the soft contribution clearly dominates
for experimentally accessible values of momentum transfer.

\section{Results on Compton scattering}

The amplitude (\ref{final}) leads to the RCS cross section  
\be
\frac{{\d} \sigma}{{\d} t} \,=\, \frac{{\d} \hat{\sigma}}{{\d} t}
                       \left [\, \frac{1}{2} (R_V^2(t) + R_A^2(t))
           -\, \frac{us}{s^2+u^2}\, (R_V^2(t)-R_A^2(t)) \,\right] \,.
\ee
It is given by the Klein-Nishina cross section 
\be
      \frac{{\d} \hat{\sigma}}{{\d} t}\, = \,\frac{2\pi\aem^2}{s^2}\; 
                         \frac{s^2+u^2}{-us}       
\ee
multiplied by a factor that describes the structure of the proton in
terms of two form factors. Evidently, if the form factors scale as
$1/t^2$, the Compton cross section would scale as $s^{-6}$ at fixed cm
scattering angle $\theta$. In view of the above discussion (see also
Fig.\ \ref{fig:FFsoft}) one therefore infers that approximate
dimensional counting rule behaviour holds in a limited range of energy.  
\begin{figure}
\parbox{\textwidth}{ \begin{center}
   \psfig{file=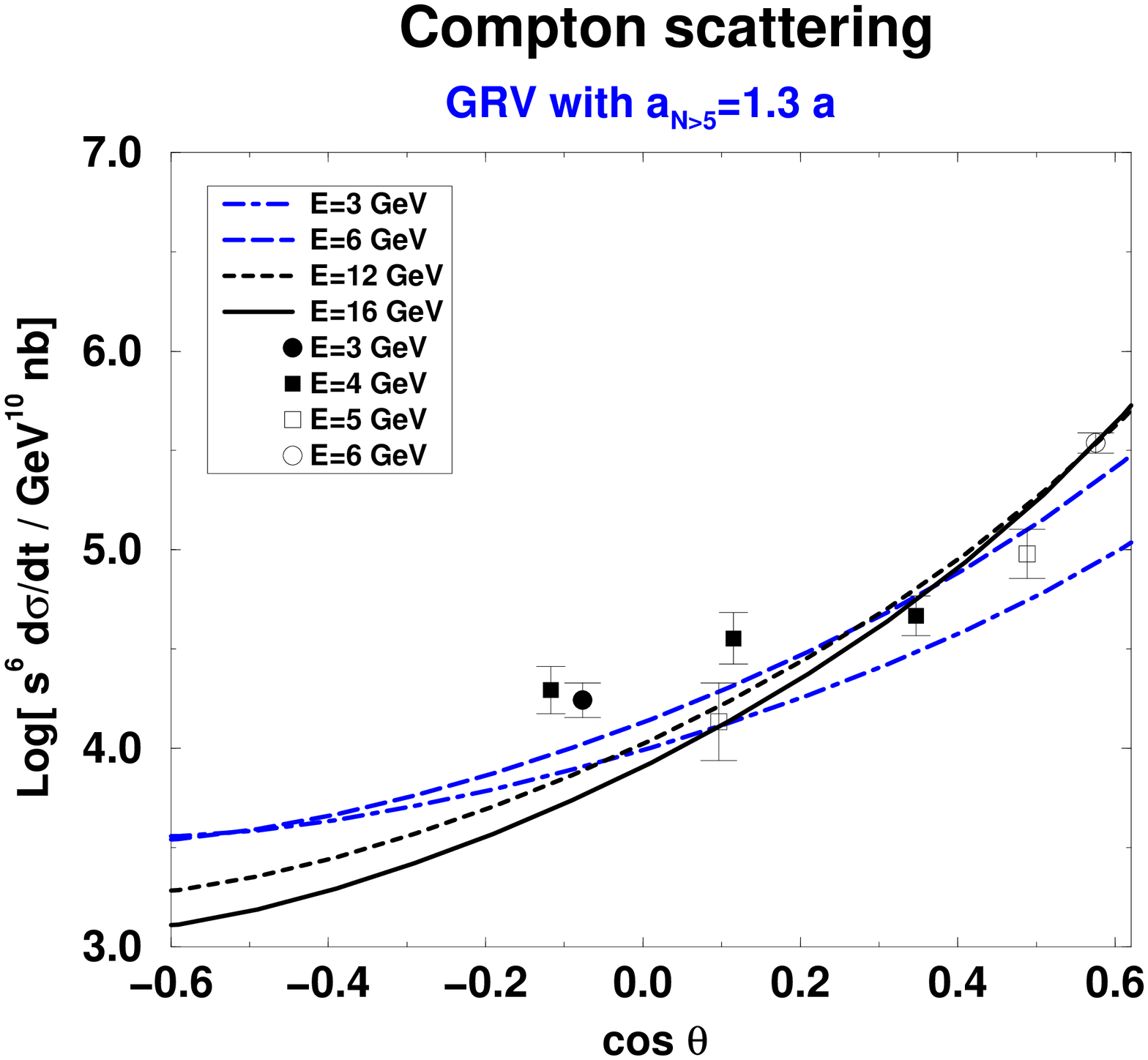, 
          bbllx=92pt,bblly=0pt,bburx=590pt,bbury=640pt,%
           width=4.5cm, angle=-90, clip=}\hspace{0.5cm}
           \psfig{file=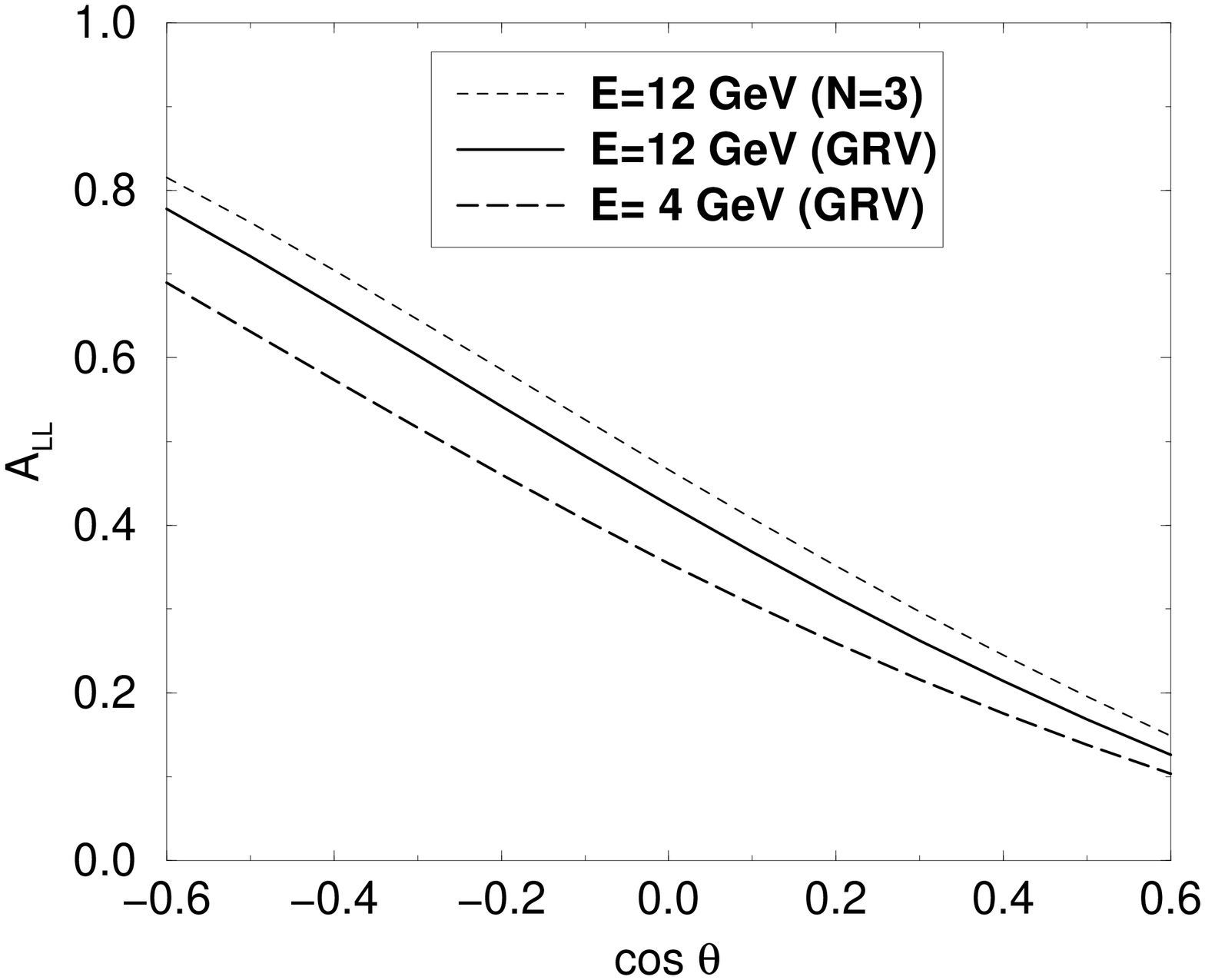, 
           width=4.5cm, angle=-90} 
\end{center}}
\caption{The Compton cross section, scaled by $s^6$, (left) 
    and the initial state helicity correlation $A_{\rm LL}$ (right)
    as predicted by the soft physics approach \protect\cite{DFJK}. Data taken
   from \protect\cite{shu79}.}
\label{fig:rcs}
\end{figure}
The magnitude of the Compton cross section is quite well predicted as
is revealed by comparison with the admittedly old data~\cite{shu79}
measured at rather low values of $s$, $-t$ and $-u$ (see Fig.\
\ref{fig:rcs}). Cross sections of similar magnitude 
have been obtained within the perturbative approach~\cite{niz91}
(evaluated from very asymmetric \das{})
and within the diquark model \cite{kro91}. The latter model is a variant
of the standard perturbative approach \cite{bro80} in which diquarks
are considered as quasi-elementary constituents of the proton \cite{ans87}. 

The \spa{} also predicts characteristic spin dependencies of the
Compton process. Of particular interest is the initial state
helicity correlation
\be
A_{\rm LL}\, \frac{{\d} \sigma}{{\d} t} = \frac{2\pi\aem^2}{s^2} \;
  R_V(t) R_A(t) \left(\frac{u}{s} - \frac{s}{u}\right) \,.
\ee
Approximately, $A_{\rm LL}$ is given by the corresponding subprocess
helicity correlation $\hat{A}_{\rm LL}=(s^2-u^2)/(s^2+u^2)$ multiplied
by the dilution factor $R_A(t)/R_V(t)$. Thus, measurements of both the
cross section and the initial state helicity correlation allows one to
isolate the two form factors $R_V$ and $R_A$
experimentally~\cite{nat99}. In Fig.\ \ref{fig:rcs} predictions for
$A_{\rm LL}$ are shown. Interestingly, the diquark model predicts the
opposite sign for $A_{\rm LL}$. For the final state helicity
correlation, $C_{\rm LL}$, and for the helicity transfer from the
incoming photon to the outgoing proton, $K_{\rm LL}$, one finds 
\be
  C_{\rm LL} = K_{\rm LL} = A_{\rm LL},
\ee
while the helicity transfer from the initial to the final state photon
, $D_{\rm LL}$, is unity since the photon helicity is conserved.
Since in the \spa{} photon helicity flips are zero and proton
helicity flips are neglected many other polarization observables are
zero, e.g.\ all single spin asymmetries or correlations between
longitudinal and sideway polarizations (normal to the particle's
momentum and in the scattering plane). In the diquark model
\cite{kro91}, on the other hand, the photon and proton helicity
flip amplitudes are non-zero although suppressed 
by either $1/\sqrt{s}$ or $1/s$; there are even perturbatively
generated phase differences between the helicity
amplitudes. Therefore, the diquark model predicts small deviations 
from zero for the latter polarization observables.

The VCS contribution to the unpolarized $ep\to ep\gamma$ cross section
is decomposed as (see e.g.\ \cite{kro96a})
\be
\frac{{\d} \sigma}{{\d}s{\d}Q^2{\d}\varphi{\d}t} \,\sim\,
           \frac{{\d}\sigma_T}{{\d}t} + 
               \varepsilon \frac{{\d}\sigma_L}{{\d}t} +
             \varepsilon \cos{2\varphi}  \frac{{\d}\sigma_{TT}}{{\d}t} 
        +  \sqrt{2\varepsilon (1+\varepsilon)} \cos{\varphi} 
                 \frac{{\d}\sigma_{LT}}{{\d}t} \,,
\label{eq:vcs}                               
\ee
where $\varepsilon$ denotes the ratio of longitudinal and transverse photon
flux in the Compton process and $\varphi$ is the azimuthal angle between
the electron and hadron planes. The partial cross sections in
(\ref{eq:vcs}) refer to the scattering of transverse 
and longitudinal photons and to the
transverse-transverse and longitudinal-transverse interference terms.   
The VCS partial cross sections have also been calculated in
Ref.~\cite{DFJK}.  
Comparing with the only other available results, namely those from the
diquark model~\cite{kro96a}, one observes that the transverse cross section
in both approaches comes out rather similar, while the other three
cross sections are generally larger and with a smoother
$Q^2$-dependence in the soft physics approach than in the diquark
model. In contrast to the diquark model the transverse-transverse
interference term, ${\d}\sigma_{TT}/{\d}t$ is strictly zero in the
limit $Q^2=0$. In addition to VCS the full $ep\to ep\gamma$ cross
section receives substantial contributions from the Bethe-Heitler
process, in which the final state photon is radiated by the
electron. Dominance of the VCS contribution requires high energies,
small values of $|\cos{\theta}|$ and an out-of-plane experiment, i.e.\
an azimuthal angle larger than about $60^\circ$. The relative
importance of the VCS and of the Bethe-Heitler contributions is rather
similar in both the approaches the \spa{}~\cite{DFJK} and the diquark model
\cite{kro96a}.

In Ref.~\cite{kro96a} the relevance of the beam asymmetry for $ep\to
ep\gamma$
\begin{equation}
A_{\rm L} = \frac{ {\d}\sigma (+) - {\d}\sigma (-)} 
                 { {\d}\sigma (+) + {\d}\sigma (-)} \,,
\end{equation}
where the labels $+$ and $-$ denote the lepton beam helicity, has been
pointed out.  It is sensitive to the imaginary part of the
longitudinal-transverse interference in the Compton process, while
${\d}\sigma_{\rm LT}/{\d}t$ measures its real part. 
In the soft physics approach, $A_{\rm L}$ is zero since all amplitudes
are real within the accuracy of the calculation. In the diquark
model~\cite{kro96a}, on the other hand, $A_{\rm L}$ is non-zero due to
the perturbatively generated phases of the VCS amplitudes. In regions
of strong interference between the Compton and the Bethe-Heitler
amplitudes the beam asymmetry is even spectacularly enhanced.
In the standard perturbative approach \cite{bro80} a
non-zero value of $A_{\rm L}$ is also to be expected.
 
\section{Scaling - evidence for pQCD?}

It is particularly interesting that the soft physics approach can
account for the experimentally observed approximate scaling, i.e\
dimensional counting rule behaviour, at least for Compton scattering
and for form factors. One may object that the perturbative explanation
(leaving aside the logarithms from the running of $\als$ and from the
evolution) works for many exclusive reactions, while in the soft
physics approach the approximate counting rule behaviour is
accidental, depending on specific properties of a given reaction. It
however seems that the approximate counting rule behaviour is an
unavoidable feature of the soft physics approach. As discussed in
Ref.\ \cite{DFJK} similar arguments as for Compton scattering lead to
a factorisation of baryon-baryon and meson-baryon amplitudes into hard
scatterings of spin 1/2 partons and soft hadronic matrix elements which are
represented by form factors analogue to the electromagnetic or Compton
ones. In elastic pp scattering, for instance, the form factor
\be
F_V^{pp}(t)\,=\, \sum_a\,  \int {\d} x\, 
         \exp{\left[\frac12 \hat a^2 t \frac{1-x}{x}\right]}  
                        \, \{ q_a(x) + \bar{q}_a(x) \} 
\label{eq:ppff}
\ee
appears. All these form factors are smooth functions of the momentum transfer
and, if scaled by $t^2$, exhibit a broad maximum in the $-t$-range
from about 5 to 15~\gev$^2$, set by the transverse hadron size, i.e.\ 
by a scale of order 1~\gev$^{-1}$. The position, $t_0$, of the maximum of $t^2
F_i$, where $F_i$ is any of the soft form factors, is determined by
the solution of the implicit equation  
\begin{equation}
- t = 4 \hat a^{-2}\, \left\langle { \frac{1-x}{x}} \right\rangle^{-1}_{F_i,t}\,.
\label{maxpos}
\end{equation} 
The mean value $\langle \frac{1-x}{x} \rangle$ comes out
around $0.5$ at $t=t_0$, hence, $t_0\simeq 8 \hat{a}^2$. Since
both sides of Eq.\ (\ref{maxpos}) increase with $-t$
the maximum of the scaled form factor, $F_i$, is quite broad. 
The scaling behaviour of the form factors lead to approximative $s^{-10}$
($s^{-8}$) scaling of ${\d}\sigma/{\d}t$ in baryon (meson)-baryon 
scattering around $s=10\gev^2$. For elastic proton-proton scattering,
shown in Fig.\ \ref{fig:pp}, fair agreement with experiment is
obtained. The proton-proton 
data \cite{ake67} show fluctuations superimposed to the $s^{-10}$
behaviour. These fluctuations, if a real dynamical feature, tell us
that there still is another momentum scale relevant in that kinematical region,
contradicting the very idea of dimensional scaling. Theoretical
interpretations of these fluctuations
have been attempted in \cite{sch75}.
\begin{figure}
\begin{center}
   \psfig{file=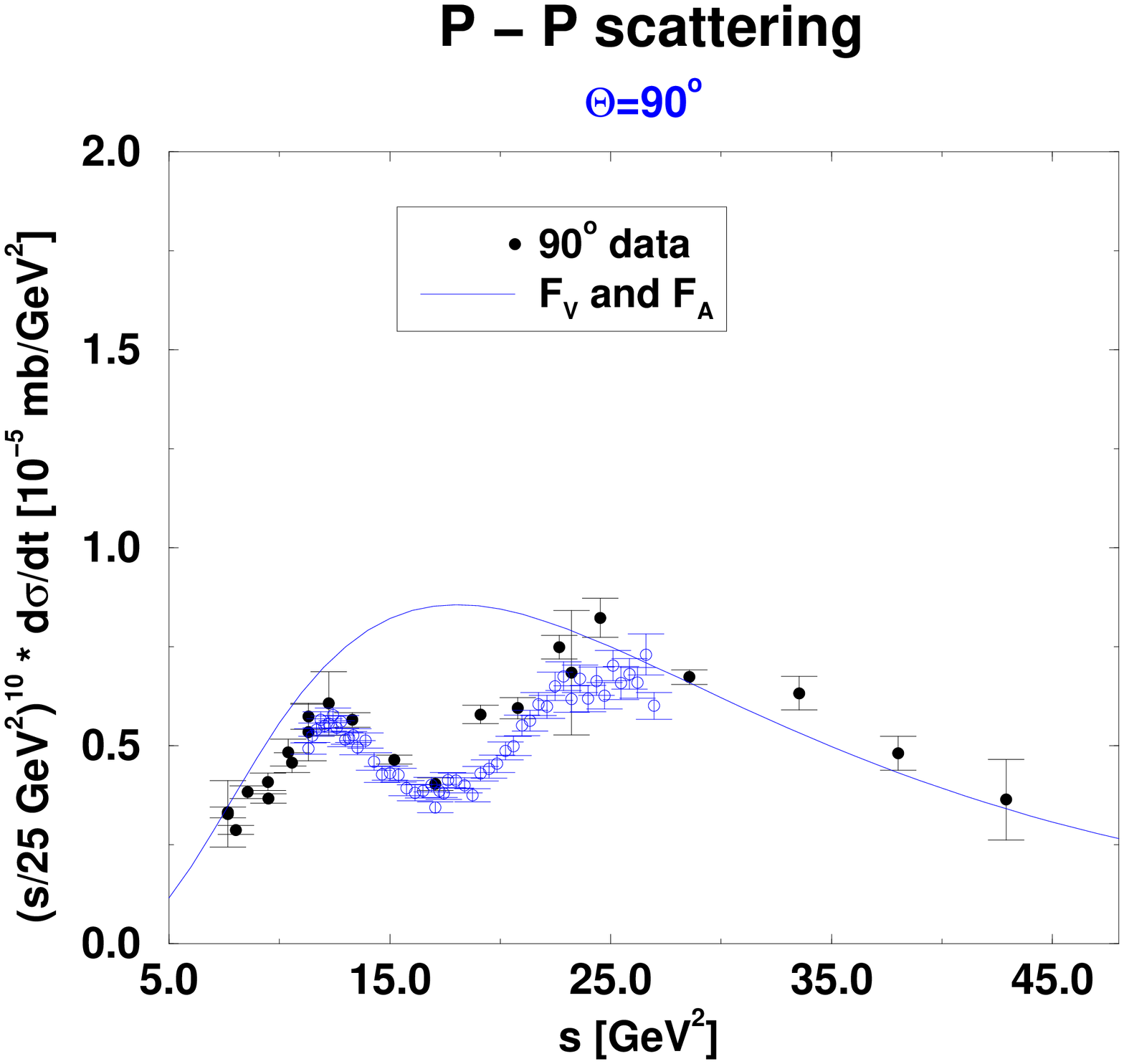, 
          bbllx=92pt,bblly=0pt,bburx=570pt,bbury=610pt,%
           width=4.5cm, angle=-90, clip=}
\end{center}
\caption{The cross section for elastic proton-proton scattering,
         scaled by $s^{10}$, at an cm scattering angle of $90^\circ$
         vs.\ $s$. The solid line is the 
         \spa{} prediction obtained in Ref.\ \protect\cite{DFJK}.
         Data taken from \protect\cite{ake67}.}
\vspace*{-0.5cm}
\label{fig:pp}
\end{figure}

\section{Summary}

The \spa{} leads to detailed predictions for RCS and VCS as well as
for form factors. The predictions exhibit interesting features and
characteristic spin dependences with marked differences to other
approaches. Dimensional counting rule behaviour for form factors,
Compton scattering and perhaps for other exlusive observables is
mimicked in a limited range of momentum transfer.
This tells us that it is premature to infer the dominance of
perturbative physics from the observed scaling behaviour . 
The soft contributions although formally representing 
power corrections to the asymptotically leading perturbative
ones, seem to dominate form factors and Compton scattering
for momentum transfers around 10 \gev$^2$. However, a
severe confrontation of this approach with accurate
large momentum transfer data on RCS and VCS is still pending.
\nopagebreak[4]
\section*{Acknowledgments}
It is a pleasure to thank Carl Carlson and Anatoly Radyushkin for
organising a most interesting meeting.
This work has been supported in part by the European TMR network ERB 4061 Pl 95 0115.

\end{document}